# Design and Implementation of a High-Accuracy Positioning System Using RTK on Smartphones

Geng Shi
Intelligent Computing and Communication Lab
Beijing University of Posts and Telecommunications
Beijing, China
e-mail: shigengbupt@163.com

Ziqiang Ying
Intelligent Computing and Communication Lab
Beijing University of Posts and Telecommunications
Beijing, China
e-mail: yingzq0116@163.com

Rongtao Xu
State Key Laboratory of Rail Traffic Control and Safety
Beijing Jiaotong University
Beijing, China
e-mail: rtxu@bjtu.edu.cn

Kan Zheng
Intelligent Computing and Communication Lab
Beijing University of Posts and Telecommunications
Beijing, China
e-mail: zkan@bupt.edu.cn

*Abstract*—In recent years, with the development of the Global Navigation Satellite System (GNSS), the satellite navigation technology has played a crucial role in smartphone navigation. To solve the problem of the low positioning accuracy in the smartphones based on GNSS, this paper proposes to apply real-time dynamic carrier phase difference technique (RTK) in the smartphones, and a real-time positioning system for smartphones based on RTK is implemented. This paper presents the implementation and experimental results of this system. This system is mainly composed of the GNSS reference station, the NTRIP system and the smartphones. The experimental results show that the system effectively improves the positioning accuracy of smartphones.

## I. INTRODUCTION

The Global Navigation Satellite System (GNSS) is a satellite-based positioning system, collaboratively providing global positioning service with high accuracy. However, the positioning accuracy, mainly determined by the GNSS signal quality and the quality of the GNSS receiver, is still a challenge to restrict the performance of GNSS. The performance of smartphone positioning is significantly affected by these factors, such as the receiver sensibility of GNSS chip [1], the ionospheric delay [2], the tropospheric delay [3], the clock bias [4], the multipath interference [5][6], and so on. Currently, improving the positioning accuracy of smartphone mainly depends on using high-quality GNSS chips. First, more efficient GNSS satellites, supporting multi-constellation service, can be provided. Second, various measurement errors can be reduced by providing multi-frequency signals, such as the GPS L5 band signal, which not only owns higher transmitted power than the L1, but also has greater bandwidth to improve the ability of jamming resistance [7].

The real-time dynamic carrier phase difference technique (RTK) is a relative positioning technology based on GNSS carrier phase measurement [8][9], where it must depend on a reference station with a known position to provide real-time corrections [10]. RTK can effectively eliminate atmospheric delay, which is the main part of the error. Therefore, the accuracy of centimeter-level can be obtained by using RTK technology [11]. Currently, smartphones that can provide raw measurement data have appeared on the market, such as Huawei Mate20. Moreover, Google has open-sourced Application Programming Interface (API) for obtaining raw measurement data from mobile phones [12], where it is possible to implement RTK on a smartphone.

To improve the positioning accuracy of smartphones, this paper proposes to apply real-time dynamic carrier phase difference technique (RTK) for positioning, and a real-time dynamic high-accuracy positioning system is designed and developed based on RTK. Besides, the system uses a satellite selection method based on the satellite elevation angle, which effectively improved the quality of the raw measurement data. The advantage of this system is that it can effectively improve the positioning accuracy of mobile phone but does not require high-performance GNSS chips. Therefore, as long as the mobile phone can provide raw measurement data, this system can improve the positioning accuracy of the mobile phone. In this work, the RTK algorithm is implemented by calling the API provided by RTKLIB, and an APP is developed to display the user's location. In addition, a GNSS reference station is constructed, and an NTRIP (Networked Transport of RTCM via Internet Protocol) system is designed to transfer GNSS data.

## II. SYSTEM ARCHITECTURE

The proposed system architecture can be shown in Fig. 1, including of three parts, i.e., reference station, NTRIP system and GNSS terminal.

### A. Reference Station

The reference station receives the GNSS signal by the GNSS antenna and then transmits the signal to the NTRIP

system, which can be acquired by the smartphone from the NTRIP system in real-time and regarded as a reference signal for the RTK solution.

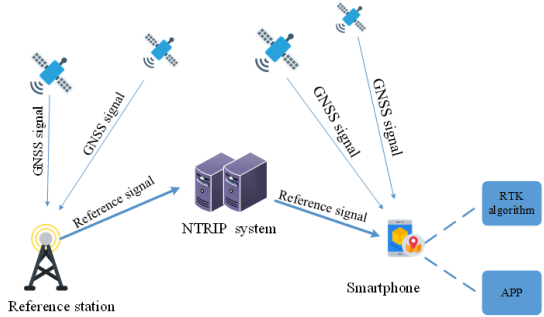

Figure 1. Illustration for system architecture.

*B. NTRIP System*

NTRIP system, including the three parts, (NtripSource, NtripServer and NTRIP Caster), is a differential data center which can be used to receive and transmit GNSS differential data. NtripSource is used to receive GNSS differential data and submit differential data to NtripServer. After users login into NTRIP Caster, they can receive the corresponding differential data. In our system, NtripSource receives the data from the GNSS reference station and sends the data to NtripServer. The smartphones can get the data from the NtripSource when it logins into the NTRIP Caster.

*C. GNSS Terminal*

In this system, the hardware platform of the GNSS terminal is based on a smartphone. The main functions of the terminal are as follows: 1) The terminal is used as a GNSS receiver to receive GNSS signals. 2) The RTK algorithm is ported to the terminal, and the terminal runs the RTK algorithm. 3) The location of the terminal is displayed in real-time on the APP developed in the system, and the APP runs on the terminal.

### III. SYSTEM IMPLEMENTATION

This section focuses on the implementation of the system, including the reference stations, the NTRIP systems, and the GNSS terminal. In this work, the hardware platform is designed and the software for this system is developed.

*A. Reference Station*

As Fig. 2 shows, the design of the reference station consists of four parts: the GNSS antenna, the data receiving module, the data processing module, and the data transmission interface.

The antenna selected in this system supports GPS, COMPASS, GLONASS, Galileo, QZSS, and SBAS systems. The GNSS receiving module is U-Blox's LEA-M8T module, which supports GPS, COMPASS, GLONASS, Galileo, QZSS, and SBAS systems. The data transmission interface adopts serial port transmission and Bluetooth transmission. The serial port transmission is realized by the chip (CH340), and the Bluetooth transmission features are implemented by the Bluetooth chip (HC-06).

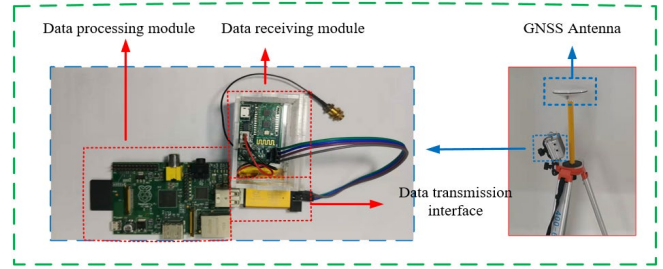

Figure 2. Reference station.

The hardware platform for data processing and transmission is based on the Raspberry Pi. It communicates with the data receiving module by UART and receives the data from the GNSS receiving module. The software for data transmission is implemented by the APIs provided in RTKLIB. By using this software, the data is uploaded to the server through the NTRIP protocol, and WIFI is used as the communication way to transmit data.

*B. NTRIP System*

In this work, an NTRIP system is implemented for receiving and transmitting GNSS data. The NTRIP is a protocol designed for streaming GNSS data to mobile receivers via the Internet. This system consists of three parts, as Fig. 3 shows: NtripClients, NtripServers and NtripCasters. NtripCaster is a GNSS differential data center for receiving and transmitting GNSS data. NtripSource is used to receive GNSS data and send data to NtripServer. NtripServer is used to submit GNSS data to NtripCaster. After NtripClient logs in to NtripCaster, NtripCaster sends GNSS data.

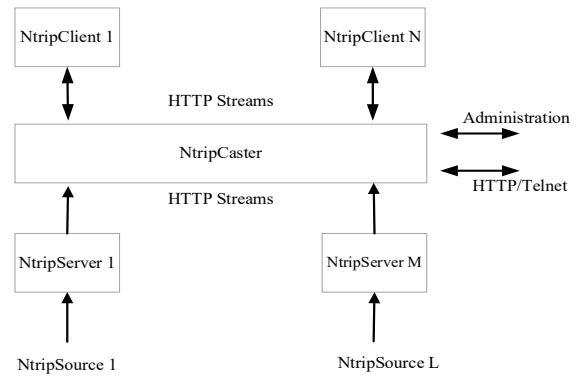

Figure 3. NTRIP system structure.

In this work, the implementation of NtripSource and NtripServer are provided by RTKLIB. The Federal Agency for Cartography and Geodesy of Germany (BKG) standard version NtripCaster program is chosen as the main source code of the NTRIP caster. The Wireless local area network (WLAN) is utilized as the communication network.

## C. GNSS Terminal

In this system, Huawei Mate20 is selected as the GNSS terminal. In this work, the software is developed for GNSS terminal. One is the software that implements the RTK algorithm, and the other is an APP that operates this system and displays the location of the terminal.

*1) Implementation of RTK algorithm:* In this work, the RTK algorithm is implemented by calling the API provided by RTKLIB. The RTKLIB is an open-source package for GNSS standards and precise positioning, consisting of a portable library and several applications to utilize the library.

In our work, the main APIs used are as follows:

*a) Strread():* The stream-read function can read raw measurement data from the input stream in a variety of ways. In this system, the implementation acquires the data from the reference station through the NTRIP, and reads the raw measurement data from the smartphone through the TCP connection.

*b) Decoderaw():* This decode-raw-data function can be used to decode received GNSS raw measurement data, which is usually encoded into a specific message format. In our work, the GNSS raw measurement data was encoded into RTCM message format, which must be decoded for future works.

*c) Rtkpos():* The RTK positioning algorithm is implemented by this API.

*d) Writesol():* This write-solution function provides the latitude and longitude to the APP, which are all displayed by the APP on the map.

Besides, the implementation of this algorithm is ported to an Android-based mobile phone. The APIs provided by RTKLIB are written in C Language, therefore the JNI technology is used to port this algorithm to smartphones.

*2) Application:* The APP is developed based on the Android, where the acquired positioning result is displayed on the map in real-time by calling the relevant APIs belonged to the Baidu map. In addition, the APP display satellite numbers and the signal carrier-to-noise ratio on the map in real-time.

## IV. DATA ACQUISITION AND PRE-PROCESSING

The GNSS raw measurement data is critical to our system. This section details the acquisition-preprocessing of raw measurement data.

### A. Data Acquisition

In this system, Huawei Mate 20 is used as a hardware to provide raw measurement data, whose GNSS receiver module supports GPS, COMPASS, GLONASS, and Galileo system.

In our work, a software tool is developed to get raw measurement data from a smartphone, which is based on Android with developed in Java language. In this work, this implementation is based on the APIs provided by Google. In recent years, Google provides the application (GNSS logger) for collecting raw measurement data in Android mobile phone. The API provides a class (GnssMeasurement) that represents GNSS satellite measurements, containing raw and calculated information, where this implementation gets the raw measurement data by calling the corresponding member function. Besides, the class GnssNavigationMessage is used to get navigation data.

By using software developed in this system for collecting data, this system can obtain raw GNSS measurement data including pseudo-range, pseudo-range rate, navigation messages, accumulated delta range or carrier, hardware (HW) clock and so on, which are used to perform RTK solution.

### B. Data Pre-processing

This section details the pre-processing of raw measurement data, where the raw measurement data must be preprocessed before it is used for RTK solution, including of two parts, first, encoding raw measurement data into RTCM message format through the encoding software, second, improving raw measurement data quality by selecting right satellites.

*1) Encoding raw measurement data:* The RTK solution module in this system supports multiple message formats, such as RINEX, RTCM, NovAtel OEM6, ubx and so on. However, the format of the raw measurement data obtained through the Google API is not supported by the RTK solution module. To satisfy the real-time requirement in our dynamic positioning system, the raw measurement data have to be encoded into the format which RTK solution can use, and the RTCM format is selected. However, there is no software tool for encoding raw measurement data into RTCM format currently. In this system, an RTCM encoding software for the raw measurement data is developed.

In this work, the RTCM 3.2 message format is selected as the encoding format, note that the message type Multiple Signal Message (MSM7) belonged to RTCM3.2 supports multi-frequency band, multi-satellite, and the high-precision operation, which is the key factor to be selected as the message type for the raw measurement data in the dynamic positioning system.

The framework of the encoding software implemented in our system is shown in Fig. 4, first, the raw measurement data is encoded into a corresponding message format according to different satellite systems. There are four kinds of RTCM3.2 message types are used in this work, such as 1077 (GPS), 1087 (GLONASS), 1097 (Galileo), and 1127 (Beidou). The encoding APIs provided by RTKLIB are used to encode the raw measurement data into the corresponding message format. What's more, the reserved fields, the message lengths, and the CRC check code are all added to the message.

*2) Improvement on the quality of raw measurement data:* The quality of the raw measurement data has a great influence on the positioning accuracy. This section focuses on ways to improve signal quality in this system. In our work, this implementation mainly improves the signal quality by selecting the right satellites. There are two reasons why our system should choose the right satellites: Excessive satellites can increase the probability of introducing low-quality signals; Due to the good geometric distribution of available Satellites, if there are more than 8 satellites, the positioning accuracy can not be better [13].

For the selection of the appropriate satellites, the geometric precision dilution (GDOP) is often used as an indicator [14][15]. The GDOP only reflects the geometric relationship between the receiver and the satellite, but the signal quality also affects the positioning accuracy. Therefore, this work considers the selection of satellite by signal quality. Owning to atmospheric delay, signal shielding, multipath, etc., the signal quality of GNSS is closely related to the elevation angle between the satellite and the receiver during GNSS signal propagation [16]. Generally, the lower the elevation angle, the greater the observed noise [17]. Therefore, the elevation angle can be used as a key parameter for selecting satellites.

To select the appropriate satellite, the data received from satellites at different angles is collected, and the pseudo-rang error is used to evaluate the quality of the data. According to the evaluation results, an appropriate angle threshold is set. This system selects the satellites with an elevation angle above the threshold.

Fig. 5 shows the distribution of all the tracked satellite. The pseudo-range error of all tracked satellite is displayed in Fig. 6. From the two graphs, it can be found that the pseudo-range measurement error of raw data received from satellite with elevation angle less than 30 degrees is greater than 8 meter. Especially the G25 satellite with an elevation angle of about 8 degrees has a pseudo-range error of up to 25 meter.

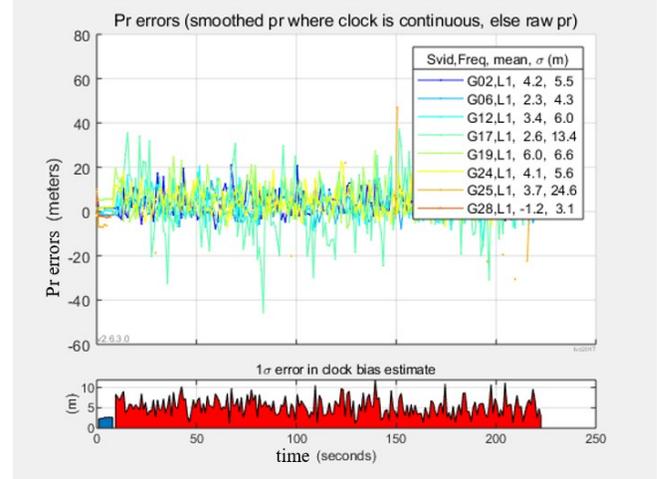

Figure 6.  Pseudo-range measurement error for tracked satellites.

According to the results, it can be concluded that in this test environment, the signal quality of satellites with an elevation angle less than 30 degrees is very poor and cannot be applied to RTK positioning. Therefore, the system sets the threshold to 30 degrees, and the satellites with an elevation angle above this threshold can be selected.

## V. SYSTEM PERFORMANC

### A. Experimental Configurations

The location of this experiment is on the campus of Beijing University of Posts and Telecommunications (see Fig. 7). The experimental environment is similar to the urban environment, and the experimental site is close to buildings. The reference station is deployed on the roof, and the distance between the reference station and the terminal is about 300 meter. Huawei mate20 was selected as the terminal device. The reference coordinates are provided by the professional RTK device (i50) which can provide centimeter-level accuracy. The GNSS signal in this experiment selects the GPS (L1) and the COMPAS (B1) signal.

In this experiment, the professional RTK device (i50) provides a reference coordinate. Then the phone is placed at the location of the reference coordinate for about 5 minutes to test, and the result of the RTK solution is saved. At the same time, this experiment takes the NMEA solution provided by the Huawei Mate20 for comparison purposes.

### B. Results and Analysis

To accurately analyze the accuracy of the proposed dynamic positioning system, this paper uses the Root Mean Square Error (RMSE) to measure the positioning accuracy.

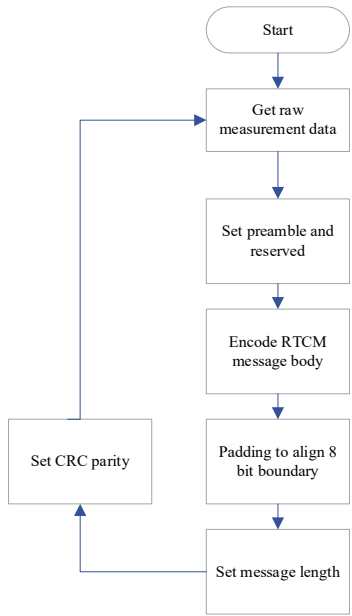

Figure 4.  RTCM encoding process.

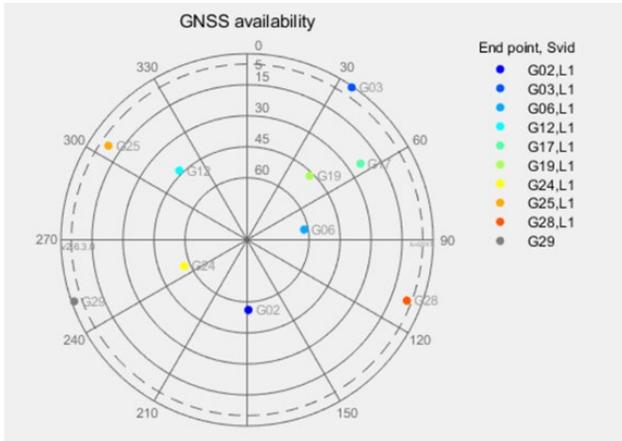

Figure 5.  Distribution of the tracked satellites.

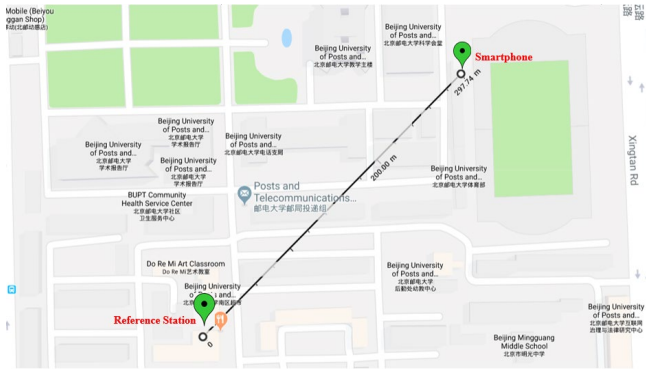

Figure 7. Map of field tests.

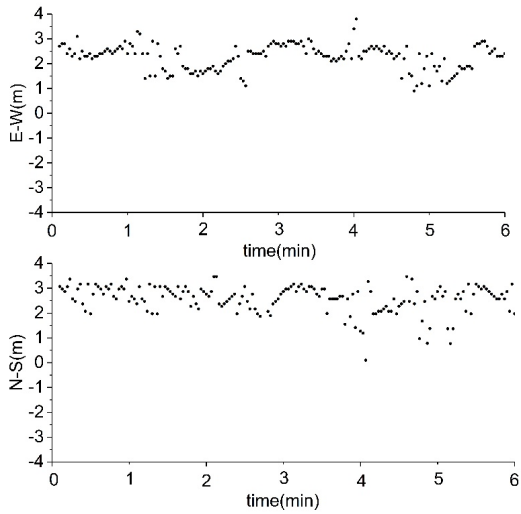

Figure 8. Error of NMEA solution.

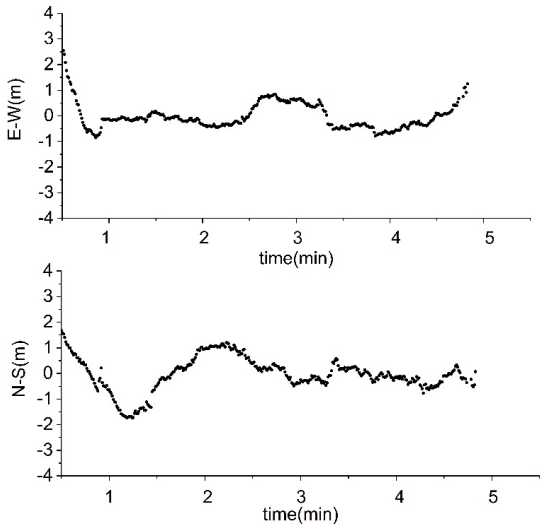

Figure 9. Error of RTK solution.

Fig. 8 shows the positioning error of the NMEA solution, and the ones of the RTK solution are displayed in Fig. 9. In the Fig. 8 and Fig. 9, The E-W represents the error of longitude, and N-S represents the error of latitude. As shown in Fig. 8, It can be found that the error of the longitude is about 2 meter, and the ones of latitude is about 3 meter. From Fig. 9, it can be found that most points of longitude or latitude are within 1 meter, but there are also a few points exceeding about 2 meter.

Besides, the corresponding RMSE of longitude and latitude is given in TABLE I. For the RTK solution, the RESM of longitude is 0.83 meter and the ones of latitude is 0.79 meter. For the NMEA solution, the RESM of longitude is 1.94 meter and the ones of latitude is 3.11 meter.

From the experimental results, it can be concluded that the RTK solution achieves greater performance compared with NMEA solution. As Table 1 shows, the RTK solution reduced the error of longitude from 1.94 meter to 0.83 meter and the one of latitude from 3.11 meter to 0.79 meter. Comparing with the results of NMEA, the ones of RTK shown in Fig. 9 quickly converge to within 1 meter, because the system uses the RTK technology to reduce the main error. However, there exist situations that the amplitudes of error are greater than 2 meter and there are two key factors.

TABLE I. RMSE OF RTK AND NMEA

| Solution | Longitude RMSE(m) | Latitude RMSE(m) |
|---|---|---|
| NMEA | 1.94 | 3.11 |
| RTK | 0.83 | 0.79 |

First, the RTK solution has a convergence process at the beginning, and there is a large deviation before the solution convergence. Second, the interruption of the satellite signal can result in significant bias in the solution results.

## VI. CONCLUSIONS

The GNSS positioning technology plays a crucial role for the positioning on smartphones. To improve positioning accuracy, this paper designs and implements a real-time positioning system based on RTK. This paper also analyzes the signal quality of satellites with different elevation angles. According to the analysis results, an effective satellite selection scheme is proposed to reduce the positioning error. To verify the proposed system, the experimental results show a great feasibility with good performance, where the positioning error of the mobile phone is less than 1 meter, indicating that the positioning accuracy of the smartphone has greatly improved. Therefore, it fully demonstrates that the system effectively promotes the positioning accuracy of smartphones.

ACKNOWLEDGEMENT

This paper is supported by the National Natural Science Funding of China (NSFC) under Grant 61671089 and National Key Technology R&D Program of China (Grant No. 2015ZX03002009).